\documentclass[]{aa}
\usepackage{txfonts}
\usepackage{graphicx}
\usepackage{natbib}
\bibpunct{(}{)}{;}{a}{}{,}

\def\kms{km\,s$^{-1}$}
\def\ms{m\,s$^{-1}$}
\def\pone{Paper~I}
\def\ione{\,{\sc i}}
\def\itwo{\,{\sc ii}}
\def\rms{rms}
\def\teff{$T_\mathrm{eff}$}
\def\logg{$\log g$}
\def\rhk{$\log R_\mathrm{HK}$}
\def\vsini{$v\sin i$}
\def\str{Str\"omgren}

\begin{document}

\title{Solar-like oscillations and 
magnetic activity of the slow rotator EK~Eri\thanks{Based on data from the HARPS spectrograph 
at the La Silla Observatory, European Southern Observatory, 
obtained under programs IDs 77.C-0080 and 78.C-0233, 
and on data from the CES spectrograph obtained from the ESO Science Archive Facility.}}

\author{
T.\ H. Dall\inst{1}
\and
H.\ Bruntt\inst{2,3}
\and
D.\ Stello\inst{3}
\and
K.\ G.\ Strassmeier\inst{4}
}

\offprints{T.\ H.\ Dall, \email{tdall@eso.org}}

\institute{
European Southern Observatory, Karl Schwarzschild Str. 2, 85748 Garching bei M\"unchen, Germany
\and
Observatoire de Paris, LESIA, 5 place Jules Janssen, 92195 Meudon Cedex, France
\and
Sydney Institute for Astronomy, School of Physics, The University of Sydney, Sydney, 2006 NSW, Australia
\and
Astrophysical Institute Potsdam (AIP), An der Sternwarte 16, 14482 Potsdam, Germany
}

\date{Received / Accepted }

\abstract
{}
{We aim to understand the interplay between non-radial oscillations and stellar magnetic activity
and test the feasibility of doing asteroseismology of magnetically active stars.
We investigate the active slow rotator EK~Eri which is the likely descendant of an Ap~star.}
{We analyze 30 years of photometric time-series data, 3 years of HARPS radial velocity monitoring, 
and 3 nights of high-cadence HARPS asteroseismic data.
We construct a high-S/N HARPS spectrum that we use to determine atmospheric parameters and chemical composition.
Spectra observed at different rotation phases are
analyzed to search for signs of temperature or abundance variations.
An upper limit on the projected rotational velocity is derived from very high-resolution CES spectra.}
{We detect oscillations in EK~Eri with a frequency of the maximum power of $\nu_\mathrm{max}=320\pm32$~$\mu$Hz, 
and we derive a peak amplitude per radial mode of $\approx0.15$~m\,s$^{-1}$,
which is a factor of $\approx 3$ lower than expected. We suggest that the magnetic field may act to suppress low-degree modes.
Individual frequencies can not be extracted from the available data.
We derive accurate atmospheric parameters, refining our previous analysis, finding 
$T_\mathrm{eff} = 5135\pm80$~K, 
$\log g = 3.39\pm0.12$,
and metallicity
[M/H] =$+0.02\pm0.04$.
Mass and radius estimates from the seismic analysis are not accurate enough to constrain the position in 
the HR diagram and the evolutionary state.
We confirm that the main light variation is due to cool spots, but that other contributions may
need to be taken into account.
We tentatively suggest that the rotation period is twice
the photometric period, i.e., $P_\mathrm{rot} = 2P_\mathrm{phot} = 617.6$~d, and that the star is a dipole-dominated oblique rotator
viewed close to equator-on. We conclude from our derived parameters that
$v\sin i < 0.40$~km\,s$^{-1}$ and we show that the value is too low to be reliably measured.
We also link the time series of direct magnetic field measurements available in the literature to our newly derived photometric 
ephemeris.
}
{}


\maketitle

\section{Introduction}
\label{intro}

\object{EK~Eri} (HR 1362, HD 27536) is a unique case of a slowly rotating ($v\sin i < 2$~\kms) 
G8 giant or sub-giant, which is over-active with respect to its 
rotation rate and evolutionary state. 
It is exhibiting brightness variations 
with a period of more than 300 days, believed to be 
due to semi-stable star spots being rotated across the projected surface \citep{strassmeier+1999}.
It has been suggested that the associated strong magnetic field is a fossil remnant from
its main sequence life as a magnetic Ap star \citep{stepien1993}.  

The star has been monitored photometrically since 1978, with first results included in \citet{strassmeier+1990}.
In a subsequent study, \citet{strassmeier+1999} 
derived a photometric period of 306.9~d from twenty years of monitoring, but noted that the light curve could
be split into two segments with different periods, 311~d (pre-1987) and 294~d (post-1992) respectively, and with a period of 
relatively small light variations in between,
possibly reflecting two distinct magnetic cycles. 
These photometric data also showed that the star gets redder when it gets fainter, which
agrees with cool spots as the cause of the light variation.  
From high-resolution spectroscopy, \citet{strassmeier+1999} also determined
the fundamental stellar parameters and showed from the radius--$v\sin i$ constraints that the star must be
seen close to equator-on, i.e. $i \approx 90^\circ$.
 
\citet{auriere+2008} published the first direct measurement of the magnetic field of EK~Eri. They find the field
to be large scale, rather than a solar-type (small-scale) field, 
dominated by a poloidal mostly axisymmetric component, resembling a dipole with a strength
of $\approx 270$~G.  They also observe some modulation over the rotation period, although their data do not cover the full
photometric period.  Their results strengthen the interpretation that EK~Eri is a descendant of a slowly rotating magnetic
Ap star which is now approaching the giant branch.

In \citet[][hereafter Paper\,I]{dall+2005} we refined the fundamental parameters of EK~Eri using new HARPS spectra and found
radial velocity (RV) variations with peak-to-peak amplitude of $\approx 100$~\ms. This variation was shown to 
correlate extremely well with the calcium H \& K activity index (\rhk) as well as with the bisector inverse
slope (BIS) of the cross-correlation function. However we found a positive correlation for the BIS rather than
the negative correlation expected from spot-induced RV variations \citep[e.g.,][]{desort+2007}.
Such correlations have previously
been attributed to fainter stellar companions contributing to the signal, effectively disguising the signal
of a planetary companion \citep[][]{santos+jvc2002,zucker+2004}.  

In this paper we present the updated results from 30 years of photometric monitoring, 
as well as from three years of RV monitoring,
and three half-nights of high-cadence RV measurements used for an asteroseismic analysis.

In Sect.~\ref{obs} we present our new observations and the data reduction. 
In Sect.~\ref{results} we present our results, which we discuss in Sect.~\ref{discussion}.
In Sect.~\ref{conclusions} we give a summary of our conclusions.

\section{Observations}
\label{obs}
\subsection{Photometry}
\label{obs_phot}
\begin{figure*}
\includegraphics[width=\linewidth]{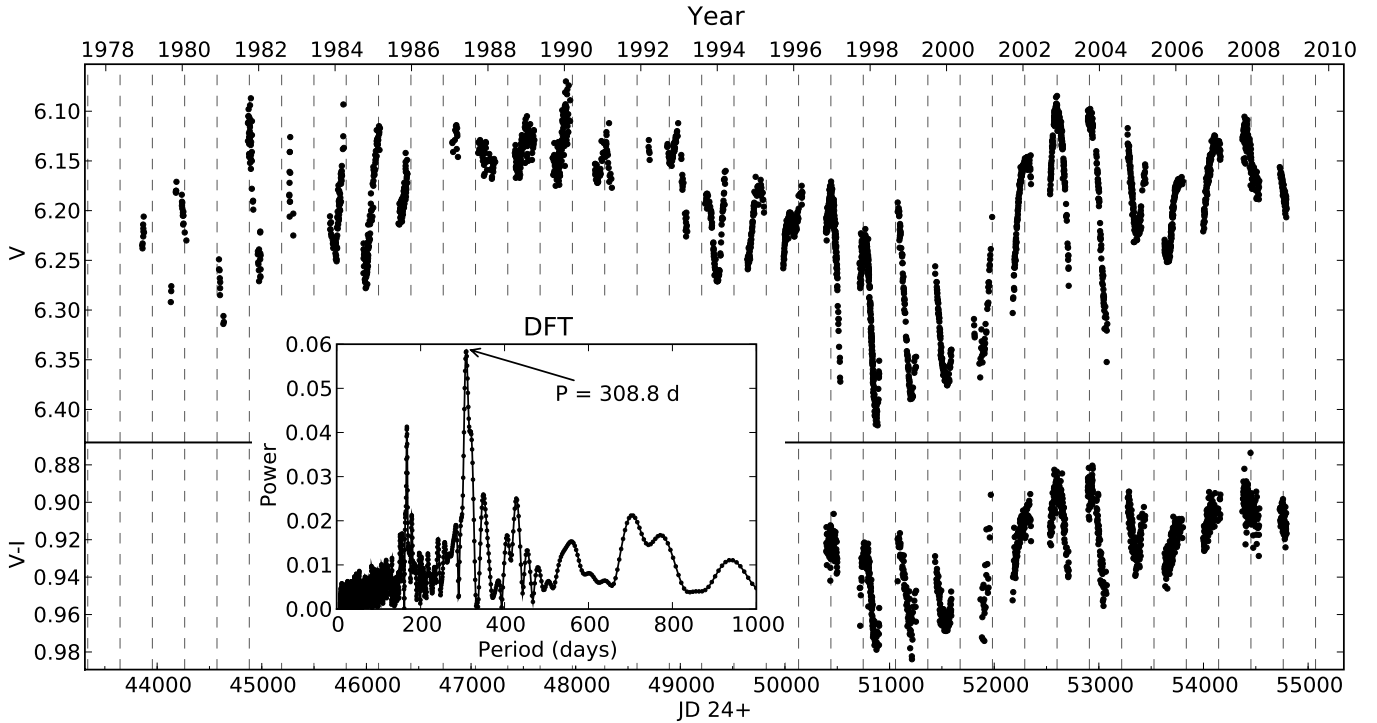}
\caption{\label{fig:tsEph}
The photometric time series in $V$ from 
1979--2009 and in $V-I$ from 1996--2009. 
The expected times of $V$ maxima are indicated with dashed lines.  
The inset shows the power spectrum from the period analysis (see Sect.~\ref{period_analysis}).
}
\end{figure*}

We present new photometric data for the years 1998 through 2009.
Photometry from 1978 through early 1998 was previously analyzed by
\citet{strassmeier+1999} and consisted of data from
various sources listed therein. The new data comes from Amadeus, one
of the two 0.75m Vienna ``Wolfgang-Amadeus'' twin automatic
photoelectric telescopes located at Fairborn Observatory in
Washington Camp in southern Arizona \citep{strassmeier+1997}.
Since late 1998, 1719 nightly $V$ and $I$ data points were acquired, each
the mean of three measurements of EK~Eri and a comparison
star. These data were transformed to the Johnson-Cousins $V(RI)_c$
system and used \object{HD~27179} as the comparison star and
\object{HD~26409} as the check star. The data quality varied between
an external rms of 4 mmag to 7 mmag in $V$. Details of the data
reduction procedure were previously given by \citet{granzer+2001}.
The main part of the data set is shown in Fig.~\ref{fig:tsEph}.

The Wolfgang telescope was used on five dedicated nights 20--24
December 2007 to monitor the star with a cadence of 3.0~min to check for 
possible high-frequency photometric variations. The
second night was lost due to poor weather, the third night was usable
for about half of the night, the others were good. 
On the three nights 20th, 23rd and 24th, the target was observed for 7.5 hours
continuously, on the 21st for 4.5 hours, and on the 22nd for 2 hours. 
The total time on target during these five nights was 29 hours of which 
the 2 hours on the 22nd were of poor quality and discarded.
A total of 542
$y$ data points were acquired. Only Str\"omgren $y$ was employed.
These data were calibrated to the standard \str\ magnitudes.
The external rms of the time-series is $\approx1.2$~mmag in $y$.

\subsection{Spectroscopy}
\label{obs_spec}
We have been conducting ultra-stable radial velocity monitoring of EK~Eri 
with HARPS \citep{harps2003,harps2004} from September 2004 to April 2007, mostly with rather uneven spacing between data points.
Typical exposure times were 480--600~s resulting in a typical peak S/N~$\approx$~300 
for individual exposures, which translates to
an intrinsic RV precision of better than 1~\ms\ per exposure. The HARPS resolution is $R=100,000$.

Accurate radial velocities are derived by the HARPS pipeline using cross-correlation with line templates (masks).
The high-precision comes from the use of a simultaneous Th-Ar calibration spectrum for accurate wavelength 
calibration and from the intrinsic high stability of the spectrograph. The resulting Cross-Correlation
Function (CCF) is fitted with a Gaussian to obtain the RV.

In March 2007 we had three consecutive half-nights of continuous monitoring
to search for oscillations and probe the asteroseismic parameters of the star. 
We used exposure times of 360--480~s which resulted in peak
S/N~$\approx$~250--300 for individual exposures at a cadence of about 8~min. 
We observed for about three hours per night,
obtaining a total of 70 spectra.

We have also retrieved archive data of EK~Eri taken with the CES high-resolution spectrograph at ESO/La Silla
Observatory. One 900~s spectrum at a central wavelength of 6155~\AA\ was obtained on UT date 2004-09-03 and one
600~s spectrum at 6463~\AA\ was obtained on UT date 2005-08-11. The spectra cover 39~\AA\ and 43~\AA, respectively.
These data were reduced using the 
CES pipeline\footnote{See {\tt http://www.eso.org/sci/facilities/\\lasilla/instruments/ces}}. The resolution is 
R~$\approx 220,000$ with S/N~$\approx500$ in both spectra.

\section{Results}
\label{results}
In the following we will investigate the high-resolution spectroscopy and long-term   
photometric time-series data. The numerical results are summarized in Table~\ref{tab:results}.

\begin{table}
 \centering
 \caption{Summary of the properties of EK~Eri.
 \label{tab:results}}
\begin{tabular}{l|r@{}l|l } \hline
Parameter            &  \multicolumn{2}{c|}{Value} & Section \\ \hline

\teff                & $5135$   & $\pm80$~K      & \ref{fund_param} \\ 
\logg                & $3.39$   & $\pm0.12$      & \ref{fund_param} \\ 
\hspace{0cm}[M/H]    &  $+0.02$ & $\pm0.04$      & \ref{fund_param} \\ 
$\xi_t$              & $1.15$   & $\pm0.03$~\kms & \ref{fund_param} \\ 
$v_\mathrm{macro}$    &  $2.3$   & $\pm0.2$~\kms  & \ref{period-radius} \\ 
$v\sin i$ (measured) & $< 1.6$ & $\pm0.4$~\kms   & \ref{period-radius} \\ 
$v\sin i$ (predicted)& \multicolumn{2}{l|}{$< 0.40$~\kms} & \ref{period-radius} \\ \hline 
$L/L_\odot$           & $14.5$   & $\pm1.6$      & \ref{mass} \\ 
$M/M_\odot$           & $1.92$   & $\pm0.13$     & \ref{mass} \\ 
$R/R_\odot$           & $4.87$   & $\pm0.29$     & \ref{mass} \\ 
Age                  &  $1.1$   & $\pm0.2$~Gyr  & \ref{mass} \\ \hline 
$P_\mathrm{rot}$      &  $617.6$ & $\pm5.0$~d    & \ref{rv-activity} \\ 
$P_\mathrm{phot}$     & $308.8$  & $\pm2.5$~d    & \ref{period_analysis} \\ 
Ephemeris            &  \multicolumn{2}{l|}{HJD$2,453,372.679 + n P_\mathrm{phot}$} & \ref{period_analysis} \\ 

\hline
\end{tabular}
\end{table}

\subsection{Photometric period and modulation}
\label{period_analysis}
\citet{strassmeier+1999} determined the period of the long-term photometric variation of 
EK~Eri based on the data obtained to that date. They also found that
the period may have changed slightly when comparing the years 1978--1986 to the years 1992--1999, possibly due to 
distinct activity cycles.  Here we derive the 
period using the full data set,
using a variety of techniques summarized in Table~\ref{tab:periods}. 
From these results we adopt a photometric period of $308.8\pm2.5$~d  (see inset of Fig.~\ref{fig:tsEph}). 
Given the errors on the individual results and the obvious spot evolution, we
cannot comfortably give the period with higher precision.
\begin{table}
\caption{\label{tab:periods}
Results from different methods of period analysis of the long-term photometry. 
The period has been found by a Gaussian fit to the peak, except for Period04 which uses a sinusoidal fit. 
The $\sigma$ is the error on the fitted peak position. 
}
\centering
\begin{tabular}{lrr}\hline
Method  & P$_\mathrm{phot}$ [d] & $\sigma$ [d] \\ \hline
Period04 & 308.99 & --- \\
DFT & 308.76 & 3.36 \\
CLEAN & 308.77 & 4.00 \\
Lomb-Scargle & 308.91 & 2.04 \\
Lafler-Kinman & 308.57 & 3.19 \\
MSL & 310.12 & 5.05 \\
PDM & 309.81 & 7.39 \\ \hline
Adopted value & 308.8 & 2.5 \\
\hline
\end{tabular}
\end{table} 
The corresponding ephemeris is given in Table~\ref{tab:results} and  
is marked by dashed vertical lines in Fig.~\ref{fig:tsEph}. 
As can be seen, the ephemeris can not always reproduce the data, 
indicating that the spots are changing size and
configuration.  From Fig.~\ref{fig:tsEph} it is 
clear that when the star is fainter, $V-I$ is higher (i.e., the star is redder)
which points to cool spots as the cause of the variation. 
Since the period is likely much longer than typical spot lifetimes we cannot hope to extract any quantitative
information on the amount of differential rotation.

We will now describe the methods used to derive the periods in Table~\ref{tab:periods}. 
Both the Discrete Fourier Transform \citep[DFT; e.g.,][]{reegen2007} 
and Period04 \citep{period04} were used, where we prewhitened
with two long periods corresponding to long-term drifts. 
Period04 is developed for asteroseismic applications where 
the light curve can be assumed to be made up of sinusoidal components. This makes it inappropriate for our 
analysis of the long term variations which are changing in amplitude and possibly also in period. The usual
formula for the period precision based on a mix of white and $1/f$ noise therefore has no physical meaning.
The CLEAN algorithm \citep[also known as iterative sine-wave fitting; e.g.,][]{frandsen+1995}
revealed only one 
main peak in the resulting amplitude spectrum after pre-whitening with the long-term drifts.
The Lomb-Scargle periodogram gives results similar to the DFT, but with slightly lower error.
The Lafler-Kinman analysis \citep{lafler+kinman1965} gives a similar result, but the periodogram 
is not as convincing --- a trait shared by the Minimum String Length (MSL) method.  The Phase Dispersion Minimization (PDM)
tend to generate higher period aliases, but even so the peak was clearly identified, although not very well
determined.

\subsection{Asteroseismic results}
\label{seismology}
\begin{figure*}
\includegraphics[width=\linewidth]{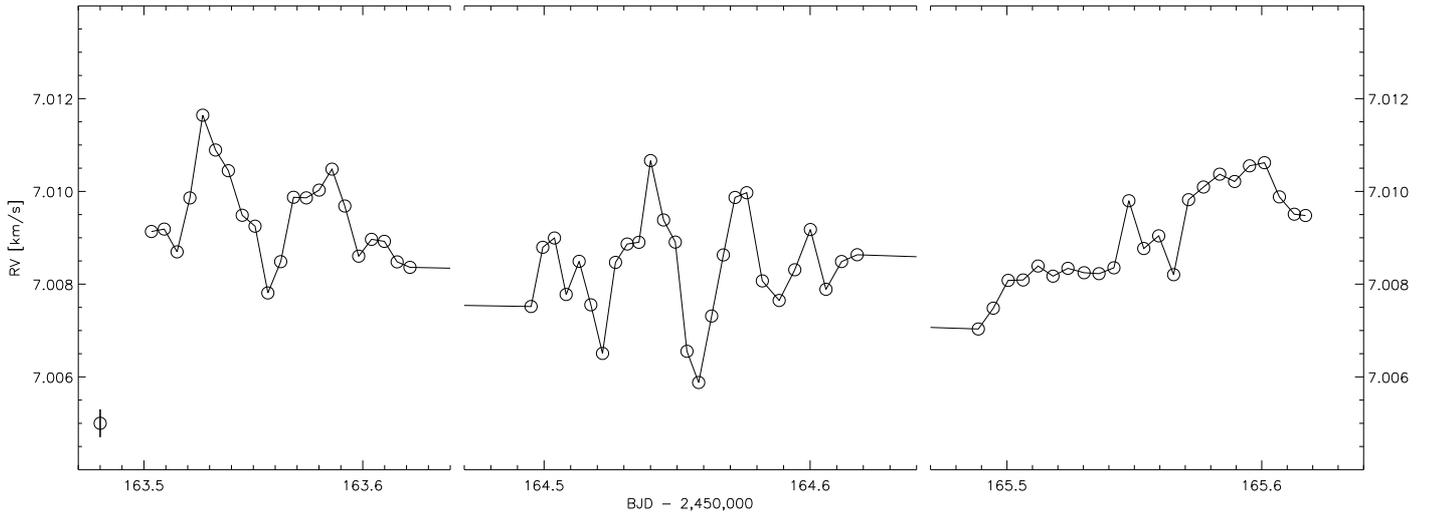}
\caption{\label{fig:seis}
RV vs.\ BJD for the three nights of high-cadence spectra used for the asteroseismic analysis. 
The mean error bar is shown at the lower left corner.
The three-night run displays clear variation, but with period variations from
night-to-night, likely due to mode beating. 
}
\end{figure*}
From its position in the HR diagram, EK~Eri is expected to show p-mode
oscillations that are stochastically excited and damped by near surface
convection similar to what is observed in the Sun and other solar-like
stars \citep{bedding+kjeldsen2003}.
We therefore carried out high-cadence ($\Delta t \approx 8\,$min) spectroscopic
monitoring during three consecutive nights using HARPS as outlined in Sect.~\ref{obs_spec}.
In addition, five nights of high-cadence photometric monitoring was conducted as described in Sect.~\ref{obs_phot}

\subsubsection{Radial velocity frequency spectrum
\label{sec:osc}}

The time series is shown in Fig.~\ref{fig:seis} and shows
significant variability of a few metres per second
(peak-to-peak) with periods of roughly 1 hour. The slight night-to-night offsets are likely due to 
slow variations in the overall activity level.
\begin{figure}
\includegraphics[width=\linewidth]{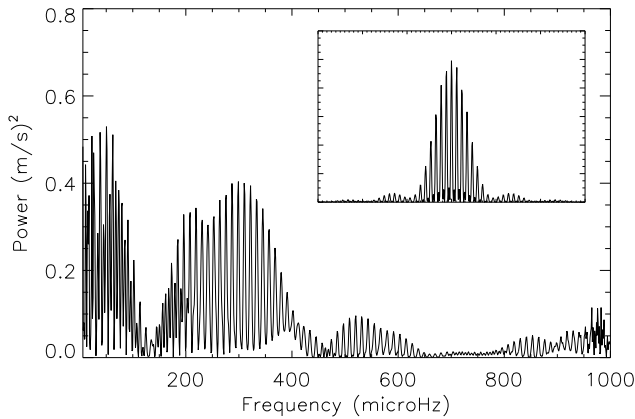}
\caption{\label{fig:fourier}
Fourier spectrum of the three nights of HARPS high-cadence radial velocity data.
The inset shows the spectral window with the same frequency scale as the main panel.}
\end{figure}
In Fig.~\ref{fig:fourier} we show the Fourier power spectrum of the time series,
which shows the variability as an excess power in the frequency range
200--400$~\mu$Hz.
This excess agrees with the expected frequency of maximum power estimated from scaling the solar value
\citep{kjeldsen+bedding1995},
which gives roughly 340~$\mu$Hz. The power at very low frequency ($\approx 50\, \mu$Hz) is predominantly caused by the slow linear trend seen on the third night and is probably not due to oscillations.

The frequency spectra of the Sun and other solar-like stars show an
almost regular series of peaks. From this, one can extract the spacing
between modes of successive radial order called the large separation,
$\Delta\nu$, which provides a very precise measure of the mean density of
the star. By scaling the solar value we find that the expected large separation
of EK~Eri \citep[using the scaling formula from][]{kjeldsen+bedding1995}
is around $20$~$\mu$Hz.

Due to the short and sparse coverage the present data set does not allow
detection of the individual frequencies or the large separation.
However, we are able to estimate the amplitude per mode from the
excess power in the Fourier spectrum using the approach by
\citet{Kjeldsen05}. First, the Fourier spectrum is converted into power
density by 
dividing by the area under the spectral window. Then we convolve the
spectrum with a Gaussian with a width of $4\Delta\nu$, to create just a
single smooth hump of excess power and finally we multiply by
$\Delta\nu/4.09$ \citep{kjeldsen+2008} and take the square root to get the
amplitude per radial mode. 

\begin{figure}
\includegraphics[width=\linewidth]{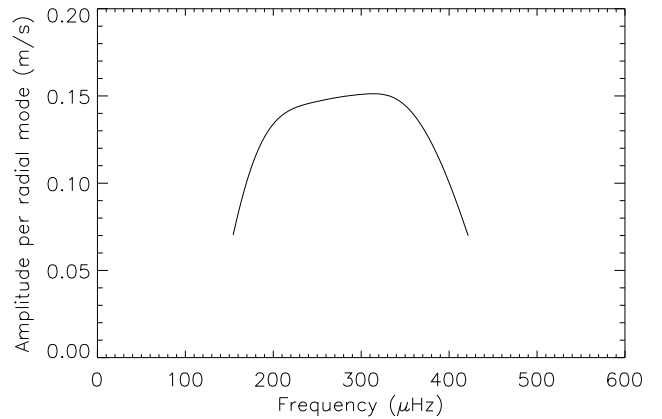}
\caption{\label{fig:amppermode}
Amplitude per radial mode. The amplitude is significantly lower than expected --- compare e.g. with Fig.~8 of \citet[][]{kjeldsen+2008}. 
See text for explanation.}
\end{figure}

The result is shown in Fig.~\ref{fig:amppermode}. The peak amplitude per
radial mode is approximately $0.15$~\ms, but with a large variation from
night-to-night, which is expected from the stochastic nature of the
excitation mechanism.  
Still, this amplitude is significantly lower than expected (by at least a factor of $\approx 3$) based on results
for stars with similar oscillation periods \citep[Fig.~8 of][]{kjeldsen+2008}. 
Part of this discrepancy
may be explained by mode beating in conjunction with the short time span of our observations.
From this analysis we find the position
of maximum power to be $\nu_{\rm max} = 320\pm32$~$\mu$Hz, where we have adopted 
a 10\% uncertainty using simulated data: for this purpose 
we use the simulator described in
\citet{Stello04}. The input mode frequencies were taken from a pulsation
model of the star derived using the ADIPLS code \citep{jcd2008b}. We
calculated the individual mode amplitudes by scaling the shape of the solar
excess
power by the acoustic cut-off frequency and normalised to the peak
amplitude found from Fig.~\ref{fig:amppermode}. The input amplitude takes
into account the 
different inertia of the modes and visibilities of whole-disk integrated observations. 
The simulations show a $\approx10$\% scatter in $\nu_{\rm max}$ and a
$\approx30$\% scatter in amplitude with assumed mode lifetimes in the range
1--20 days.
In addition, we tried to utilise the simulations to obtain a mode lifetime,
but concluded that the data did not allow a robust estimate.

\begin{figure*}
\resizebox{\hsize}{!}{\rotatebox{90}{\includegraphics{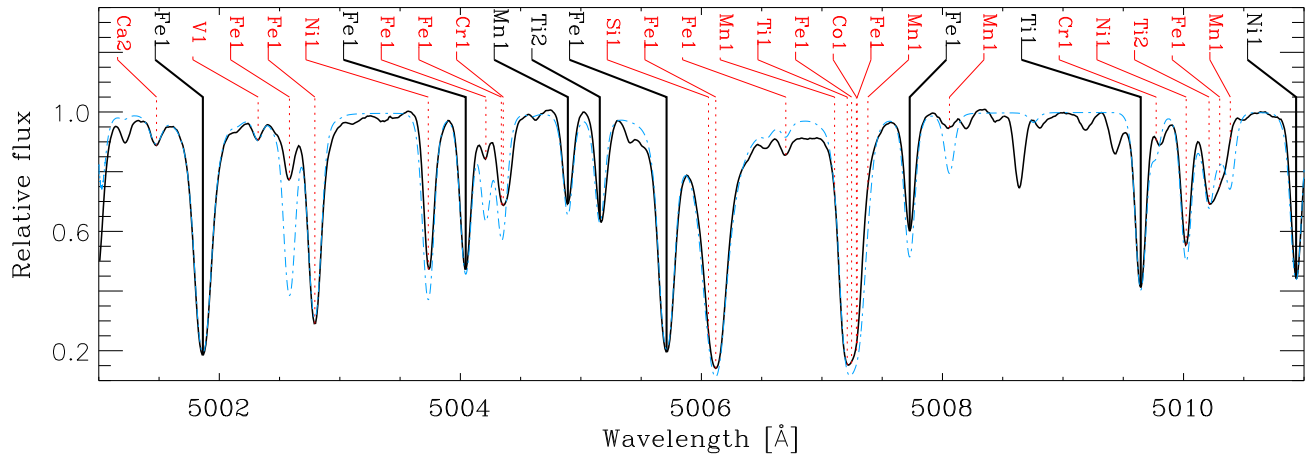}}}
\caption{\label{fig:vwaview}
A section of the combined HARPS spectrum illustrating the high S/N. 
The observed and synthetic spectra are shown with 
solid and dot-dashed lines, respectively.
The main lines contributing to the spectrum are marked by vertical lines. 
The solid vertical lines mark the least blended lines that are used in the abundance analysis.
}
\end{figure*}

\subsubsection{Short-term photometric variations}
We also searched for variability due to oscillations in the high-cadence photometric data.
Of the three full nights obtained, a period analysis reveals no periodicity on 
two of the nights, while on a third night a single peak with a false alarm 
probability of 9.4\% was detected at $f=26.016$~d$^{-1}$ ($301.11$~$\mu$Hz) 
or around 55~min at an amplitude of $\approx 5$~mmag. 
This frequency corresponds well to the envelope found from the spectroscopic time series.
However, based on the scaling relations of \citet{kjeldsen+bedding1995} 
and the observed RV amplitude, the photometric amplitude of EK~Eri is 
expected to be $\approx 50$~ppm, which is a factor of 100 lower than
the observed peak. Although this peak appears significant, we believe it to be an artifact.

\subsection{Effective temperature, gravity, abundances}
\label{fund_param}

To determine the fundamental parameters 
of EK~Eri we analysed a co-added spectrum with high S/N.
This spectrum was made from the three consecutive nights of asteroseismic observations.
We shifted the HARPS spectra by 
the individually measured velocity shifts and calculated the sum.
The blue arm of the spectrum from 4402--5304\,\AA\ has 
S/N of 700 and the red arm from 5337--6905\,\AA\ has
S/N of 1,300 in the continuum. 
We made a careful normalization
by identifying continuum points in a synthetic
spectrum with the same parameters as EK~Eri.
A small part of the co-added spectrum is shown in Fig.~\ref{fig:vwaview}. 
Also plotted is a computed synthetic spectrum which shows some discrepancies,
likely due to missing lines (e.g.\ 5008.7\,\AA) 
or erroneous $\log gf$ values (e.g.\ 5002.6\,\AA).

We analysed the normalized spectrum using the
software package VWA \citep{bruntt+2004, bruntt+2008}. 
The software uses atmospheric models interpolated 
in the grid by \citet{heiter+2002} and 
atomic data from the VALD database \citep{kupka+1999}.
The computation of abundances relies on 
synthetic spectra calculated with the SYNTH code \citep{valenti+1996}.

We used VWA to automatically select 1019 of the least blended lines.
However, as part of the analysis we corrected the oscillator 
strengths ($\log gf$) by measuring abundances for 
the same lines found in a spectrum of the Sun 
from the Atlas by \citet{hinkle+2000} (originally from \citealt{kurucz1984}): 
627 lines were found in common with the Sun and could be corrected,
and only these lines were considered in the further analysis.
This differential analysis
leads to a significant improvement in the \rms\ scatter 
of the abundances of Fe\ione\ and Fe\itwo\ from 0.12 to 0.06 dex. 
For other elements the improvement in the scatter is typically 30\%.

The fundamental atmospheric parameters of EK~Eri were
determined by adjusting the microturbulence ($\xi_t$), \teff, and \logg\
until no correlation was found between the abundances
of Fe\ione\ and their equivalent width or the excitation
potential and with the additional requirement that the mean Fe abundance
is the same found from neutral and ionized lines.
The resulting parameters are $T_{\rm eff}=5135\pm60$\,K,
$\log g = 3.39\pm0.06$ and $\xi_t = 1.15\pm0.05$\,\kms.
We estimated the uncertainties by perturbing 
$\xi_t$, \teff\, and \logg\
in a grid around the derived values by
$\pm 0.2$\,\kms, $\pm 200$~K, and $\pm0.2$\,dex, respectively,  
and measuring the change in the correlations described 
above (see \citealt{bruntt+2008}). These uncertainties are
``internal'' errors since we assume the model atmosphere describe
the actual star. We have added an addition systematic
error of 50~K and 0.1~dex to \teff\ and \logg\ in Table~\ref{tab:results}.

\begin{figure}
\resizebox{\hsize}{!}{\rotatebox{90}{\includegraphics{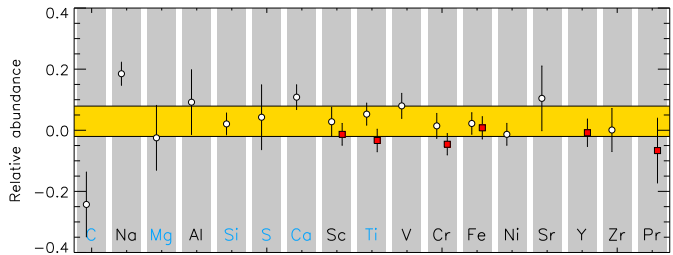}}}
\caption{\label{fig:vwares} The abundance pattern of EK~Eri measured relative to the Sun.
Circles are mean abundances of neutral elements and squares are for singly ionized species.}
\end{figure}

The abundance pattern for $17$ elements is shown in Fig.~\ref{fig:vwares}
and listed in Table~\ref{tab:ab}. 
The uncertainties on the abundances are the quadratic sum 
of the standard deviation of the mean value
and the contribution due to the uncertainty on 
the model parameters ($\sigma_{\rm model} = 0.035$~dex).
Our estimate of the metallicity is based on the mean
abundance of iron-peak elements with at least 
$10$ lines (Si, Ti, V, Cr, Fe and Ni)
giving ${\rm [M/H]} = +0.02\pm0.04$ for the combined spectrum.

From Fig.~\ref{fig:vwares} we do not see evidence for peculiarities 
in the abundances, which might otherwise have hinted at an earlier phase 
as a chemically peculiar star. 
Any abundance peculiarities present initially in the photosphere
seem to have been efficiently homogenized with the deeper layers
as the star evolved off the main sequence.

Our results on the fundamental parameters
and abundances are different from what we found in \pone\ at a level 
that may seem incompatible with the uncertainties: 
\teff\ is 100~K cooler and \logg\ is 0.1 dex lower in our new analysis.
The quality of the spectrum used in \pone\ was lower 
and therefore only half as many lines could be used
(${\rm S/N} \simeq 500$ near 6300\,\AA\ in \pone\ 
compared to $1300$ in our new spectrum).
Both analyses were done by measuring abundances differentially 
to the same lines in a spectrum of the Sun. In \pone\ we 
used a sky spectrum as a proxy for the Sun, while in the current analysis we use the 
high-quality spectrum from \cite{kurucz1984}. 
This difference could introduce a systematic offset.
It can be debated whether it is valid to use
the differential approach since EK~Eri is a more massive and evolved star
compared to the Sun. In fact, the differences we find in 
\teff\ and \logg\ may reflect realistic uncertainties to be adopted for the star.

A more intriguing idea is that the lower \teff\ we find is due to
a higher amount of spot coverage at the time of observation.
To investigate the evidence for changes in \teff\
at different rotational epochs, we have constructed spectra
by combining subsets of our new data. 
These four spectra, designated E1 to E4,
are listed in Table~\ref{tab:epochs} along with the parameters 
determined from them. The time intervals corresponding to these spectra are 
indicated in Fig.~\ref{fig:tsAll}. 
Firstly, we do not see any significant abundance variation in the E1--E4 spectra. 
Secondly, $T_{\rm eff}$ seems to be systematically lower 
when the star is faint (which would be expected if it would be an effect of changes in spot coverage).
However, we do not claim that the slight differences between the E1{\ldots}E4 spectra
are real, since the variation of \teff\ and \logg\ are only 40~K and 0.1~dex respectively,
which are well within the uncertainties of the main analysis.
Although the relative change in \teff\ can be measured very precisely,
we must consider
that the model atmospheres have been interpolated 
from the relatively coarse grid of \cite{heiter+2002} which
has a finite grid step size of 200~K in \teff\ and 0.2~dex in \logg.
This is the main argument that we do not claim 
the slight differences in \teff\ to be a physical phenomenon.

\begin{table}
 \centering
 \caption{Abundances of 17 elements in EK~Eri.
The columns contain the element name and ionization stage,
the mean abundance relative to the Sun, and the number of spectral lines used.
 \label{tab:ab}}
\begin{tabular}{l|lr||l|lr}
El. & \multicolumn{1}{c}{$A$}  & $N$ & El. & \multicolumn{1}{c}{$A$}  & $N$ \\ 
\hline
  {C  \sc   i} & $ -0.24          $  &   2  &   {V  \sc   i} & $ +0.08  \pm0.04 $  &  15  \\  
  {Na \sc   i} & $ +0.19  \pm0.04 $  &   3  &   {Cr \sc   i} & $ +0.01  \pm0.04 $  &  20  \\ 
  {Mg \sc   i} & $ -0.02          $  &   2  &   {Cr \sc  ii} & $ -0.05  \pm0.04 $  &   4  \\ 
  {Al \sc   i} & $ +0.09          $  &   2  &   {Fe \sc   i} & $ +0.02  \pm0.04 $  & 300  \\ 
  {Si \sc   i} & $ +0.02  \pm0.04 $  &  31  &   {Fe \sc  ii} & $ +0.01  \pm0.04 $  &  15  \\ 
  {S  \sc   i} & $ +0.04          $  &   1  &   {Ni \sc   i} & $ -0.01  \pm0.04 $  &  63  \\ 
  {Ca \sc   i} & $ +0.11  \pm0.04 $  &   8  &   {Sr \sc   i} & $ +0.10          $  &   1  \\ 
  {Sc \sc   i} & $ +0.03  \pm0.05 $  &   5  &   {Y  \sc  ii} & $ -0.01  \pm0.04 $  &   5  \\ 
  {Sc \sc  ii} & $ -0.01  \pm0.04 $  &   4  &   {Zr \sc   i} & $ +0.00  \pm0.07 $  &   3  \\ 
  {Ti \sc   i} & $ +0.05  \pm0.04 $  &  50  &   {Pr \sc  ii} & $ -0.07          $  &   1  \\ 
  {Ti \sc  ii} & $ -0.03  \pm0.04 $  &  11  &                &                     &      \\
\end{tabular}
\end{table}

\begin{table*}
\caption{\label{tab:epochs}
Results of the spectroscopic analysis using sub-sets E1--E4 of the HARPS spectra.
$N$ is the number of spectra summed and S/N is for the red part of the spectrum. 
The photometric phase $\phi_\mathrm{phot}$ is relative to the ephemeris zero point given
in Table~\ref{tab:results}. The phases are marked in Fig.~\ref{fig:rvbjd}.
The results from the combined spectrum and from Paper~I are listed for convenience.
}
\centering
\begin{tabular}{llcrrrrr}\hline
   & \multicolumn{1}{c}{Time span} & 
                         $\phi_\mathrm{phot}$ &$N$  & S/N   & \teff\ [K] & \logg   &  [Fe/H]  \\ \hline
E1   & 2006-02-11 --- 2006-04-02 & 0.31 -- 0.47 &  5  & 700   & $5150$     & $3.34$  &  $+0.06$ \\
E2   & 2006-07-17 --- 2006-09-08 & 0.82 -- 0.99 &  7  & 900   & $5125$     & $3.44$  &  $+0.09$ \\
E3   & 2006-12-28 --- 2007-01-03 & 0.35 -- 0.37 &  5  & 530   & $5165$     & $3.37$  &  $+0.04$ \\
E4   & 2007-03-29 --- 2007-04-02 & 0.64 -- 0.65 &  4  & 410   & $5135$     & $3.42$  &  $+0.06$ \\ \hline
Comb.& 2007-03-03 --- 2007-03-05 & 0.56         & 70  & 1,300 & $5135$     & $3.39$  &  $+0.02$ \\ \hline
Paper~I & 2004-10-01 --- 2005-03-17 &          &   6  & 500   & $5240$     & $3.55$  &  $+0.09$ \\ \hline
\end{tabular}
\end{table*}

\begin{figure}
\includegraphics[width=0.9\linewidth]{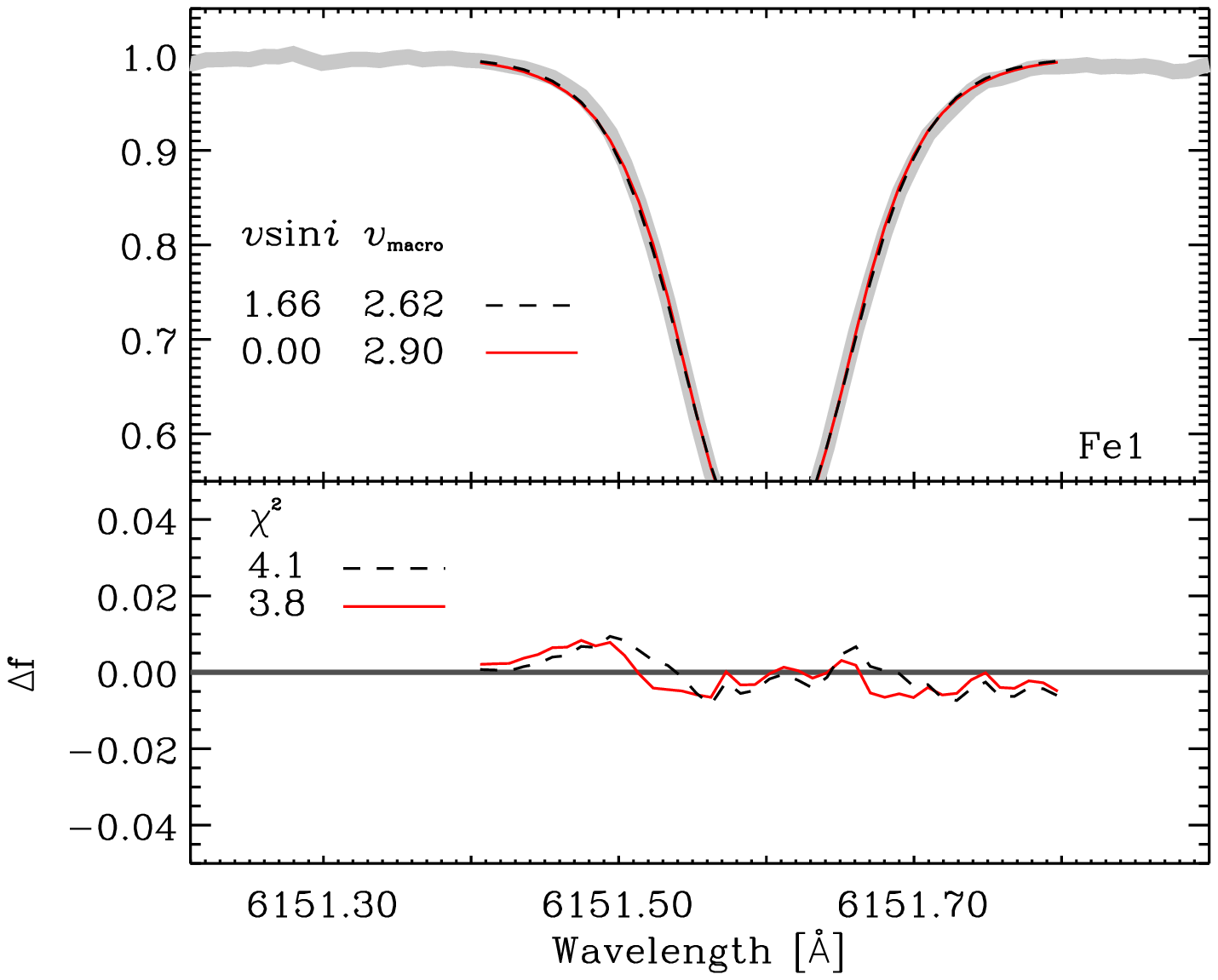}
\includegraphics[width=0.9\linewidth]{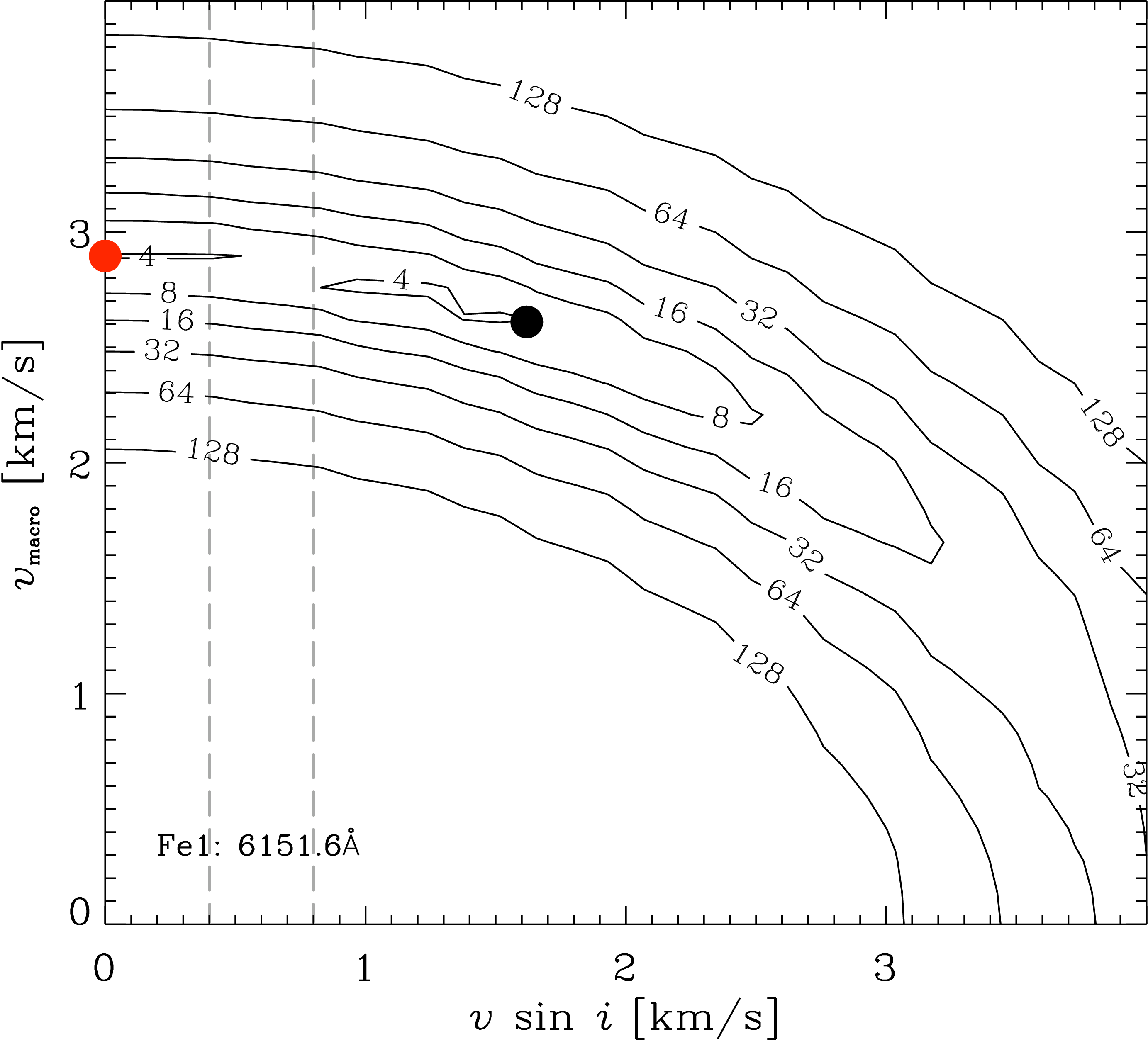}
\caption{\label{fig:vsini}
Synthetic profiles of the \ion{Fe}{i} 6151.6~\AA\ line fitted to the CES spectrum.
The Upper panel shows the observed spectrum in grey, and two fits using
the values of $v\sin i$ and $v_\mathrm{macro}$ listed. 
The residuals of the fits are plotted underneath. 
The lower panel shows the $\chi^2$ contours as a function of 
$v\sin i$ and $v_\mathrm{macro}$ for the same Fe line 
with the position of the two plotted fits marked by filled circles (red and black).
Indicated with dashed vertical lines are the limits on $v\sin i$ imposed 
by the stellar radius and the $P_\mathrm{rot} = 2 P_\mathrm{phot}$ and 
$P_\mathrm{rot} = P_\mathrm{phot}$ alternatives.
}
\end{figure}

\subsection{Mass estimates}
\label{mass}
To estimate the mass of the star we followed the approach of \cite{Stello08}
by applying a scaling relation of $\nu_{\rm max}$, 
which includes the luminosity, effective temperature, and mass.
To estimate the luminosity we use the 
$V$ magnitude, bolometric correction and the parallax. 
The magnitude $V\simeq6.0$ is the brightest measured magnitude, hence
we assume it corresponds to no or low spot activity.
This value is probably only accurate to within $\pm0.1$~mag.
We adopted a bolometric correction from the tables in \cite{bessell+1998} 
for the \teff\ and $\log g$ obtained from the spectroscopic analysis.
Finally, we use the parallax $16.31\pm0.47$~mas \citep{vanLeeuwen2007} to get
$L/L_\odot=14.5\pm1.6$ and $M_\mathrm{bol}=1.952\pm0.087$. 
To obtain the mass we use the observed value of $\nu_{\rm max}=320\pm32$~$\mu$Hz,
and get $M/M_\odot = 2.1\pm0.3$.

Another way to estimate the mass is to compare the location of the star
in the HR diagram with theoretical isochrones
or evolutionary tracks.  \cite{Stello2009} 
compared different methods to estimate 
the fundamental parameters of stars from basic photometric indices. We
adopted their SHOTGUN approach, which uses BASTI isochrones
\citep{Pietrinferni04} and models without overshoot, 
taking as input the metallicity, effective temperature,
and luminosity.
The output from the program is the mass, radius and age of the star:
$M/M_\odot = 1.92\pm0.13$, $R/R_\odot = 4.87\pm0.29$, and age $1.1\pm0.2$~Gyr. 
From $M$ and $R$ we get  $\log g = 3.35\pm0.06$, 
which is very similar to the value $\log g = 3.39\pm0.06$
from the spectral analysis in Sect.~\ref{fund_param}. 
Including overshoot in the BASTI models results in only slightly different values;
$M/M_\odot = 1.80\pm0.10$, $R/R_\odot = 4.84\pm0.35$, age $1.5\pm0.2$~Gyr, and $\log g = 3.33\pm0.07$. 
In Table~\ref{tab:results} we list the results using the standard BASTI isochrones without overshoot.

These two independent mass estimates are in rough agreement.  
A seismologically determined mass estimate is potentially more accurate than isochrone fitting
if one can measure the large separation $\Delta\nu$ or
the position of maximum power $\nu_{\rm max}$ \citep{basu+2009}. Since our data do not allow
us to measure $\Delta\nu$  and the error on $\nu_{\rm max}$ is rather large, 
we adopt the mass estimate based on the SHOTGUN approach.

\section{Discussion}
\label{discussion}

\subsection{Rotation period, $v\sin i$, and radius}
\label{period-radius}
The rotational velocity of EK~Eri is notoriously hard to measure due the the very slow rotation. The first accurate determination
was done by \citet{strassmeier+1999} from $R=120,000$ spectra, from which they derived $v\sin i = 1.5 \pm 0.5$~\kms\ with 
an adopted value of the macroturbulent 
velocity $v_\mathrm{macro}=5.0$~\kms, with a Fourier analysis of the line profile as a check of the validity of the
adopted $v_\mathrm{macro}$.  
In Paper~I we derived $v\sin i=1.0\pm 0.5$~\kms\ by simple profile fitting to 1D atmospheric model spectra,
assuming a similar value of the macroturbulence.

As we will argue in Sec.~\ref{rv-activity}, we suggest that the rotational period of EK~Eri
could be twice the photometric period, i.e. close to 617~d.
We have derived $4.87\pm0.29$~R$_\odot$ (see Sect.~\ref{seismology}), 
which is in good agreement with the value of $R = 4.78$~R$_\odot$ we derive 
using Eq.~1 from \citet{strassmeier+1999}. 
Adopting $P_\mathrm{rot} = 2 P_\mathrm{phot}$, this means that the radius of the star will
be $R/R_\odot = 12.2\, v\sin i$, where we assume that $i \simeq 90^\circ$.  With the derived radius of $4.87$~R$_\odot$, 
this translates to an expectation for the measurement of $v\sin i$, namely $v\sin i < 0.40$~\kms.  
If on the other hand $P_\mathrm{rot} = P_\mathrm{phot}$, then $v\sin i$ must be less than $0.80$~\kms, 
as already noted by \citet{strassmeier+1999}, 
and this value can thus be regarded as a safe upper limit on $v\sin i$.

Using the high-resolution CES data, and our newly derived stellar parameters, we have attempted to derive $v\sin i$ and  
$v_\mathrm{macro}$.
For this analysis we chose 7 
lines from the two CES spectra, chosen to be non-blended based on the line list
obtained from VALD \citep{kupka+1999} and visual evaluation of the line symmetry. 
The synthetic spectrum is broadened with
different values of $v\sin i$ and $v_\mathrm{macro}$ in a fine grid in the range 0--4~\kms\ for both parameters, 
and the $\chi^2$ is calculated.
At values of $v_\mathrm{macro} > 4$~\kms\ the fits start to deviate significantly and we thus note that a
value of $v_\mathrm{macro,RT} = 5$~\kms\ \citep[originally from][as a ``typical'' value]{gray1992} is inconsistent with our results.

There is a strong correlation between the broadening caused by rotation and by macroturbulence which
is very difficult to separate even though their exact forms are slightly different.
In general we find a continuum of solutions, as illustrated in 
Fig.~\ref{fig:vsini}, which shows the reduced  $\chi^2$ surface for
one spectral line for different values of $v\sin i$ and  $v_\mathrm{macro}$.
The $\chi^2$ surfaces for the other lines look very similar.
We find that \vsini\ is so low that it cannot be determined reliably and only
upper limits can be set. 
Note that we do not include Zeeman broadening, which in any case would contribute to an even 
lower limit on $v\sin i$.

\subsection{Radial velocity and activity variations}
\label{rv-activity}
\begin{figure}
\includegraphics[width=\linewidth]{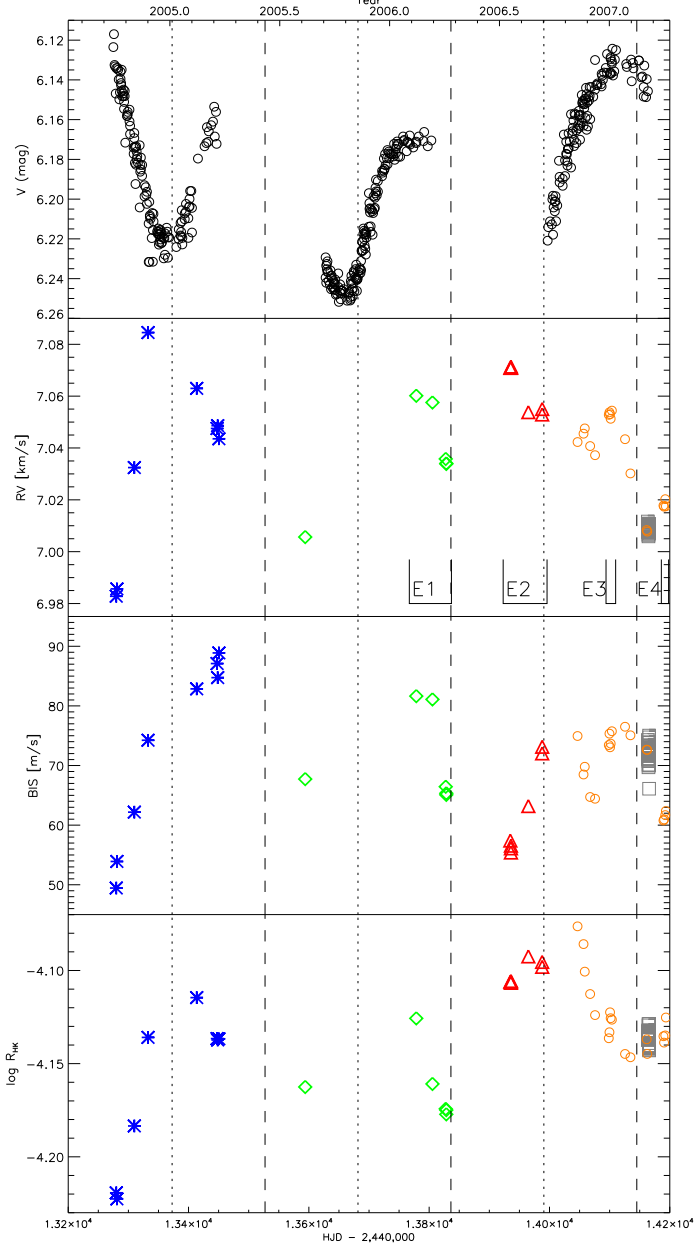}
\hskip 0.2cm
\caption{\label{fig:tsAll}\label{fig:rvbjd}
Comparison of the photometric and spectroscopic time series 
over the 3 years of simultaneous monitoring.
The colors and the symbols in the 3 lower plots are: 
blue ($\ast$): 2004-09-30 --- 2005-03-20, 
green ($\diamond$): 2005-08-10 --- 2006-04-02, 
red ($\triangle$): 2006-07-17 --- 2006-09-08, 
orange ($\circ$): 2006-11-06 --- 2007-04-02 (except the three-night run in March 2007),
and grey ($\Box$, at $\sim$2007.4): 2007-03-03 --- 05 (3 nights of high-cadence data, cf.\ Fig.~\ref{fig:seis}). 
The expected times of photometric minima (vertical dotted lines) and maxima (vertical dashed lines) are indicated.
The data segments E1--E4 are marked (see Table~\ref{tab:epochs} and Sect.~\ref{fund_param}).
}
\end{figure}

\begin{figure}
\includegraphics[width=\linewidth]{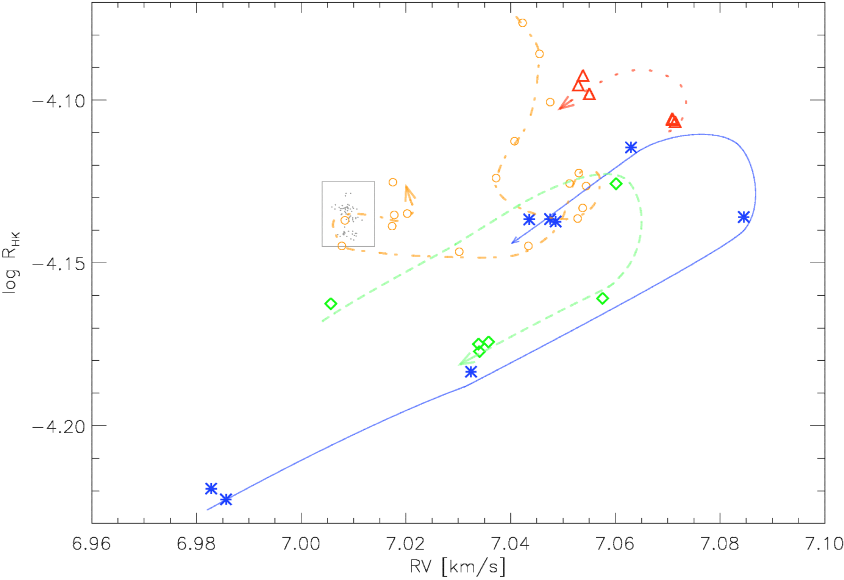}
\caption{\label{fig:rhk} Activity index $\log R_\mathrm{HK}$ vs.\ RV 
using the same symbols as in Fig.~\ref{fig:rvbjd}. 
The lines and arrows indicate the evolution with time.
The small grey points inside the box are the 3 nights of high-cadence data.
The change in activity level during these nights 
is only about 0.02 in $\log R_\mathrm{HK}$ 
which is much smaller than the long-term variations.
The clear correlation between activity level and RV
shows that the majority of the RV variations come from activity-related causes. 
The overall level seems to have increased between 2004-05 and 2006-07. }
\end{figure}

\begin{figure}
\includegraphics[width=\linewidth]{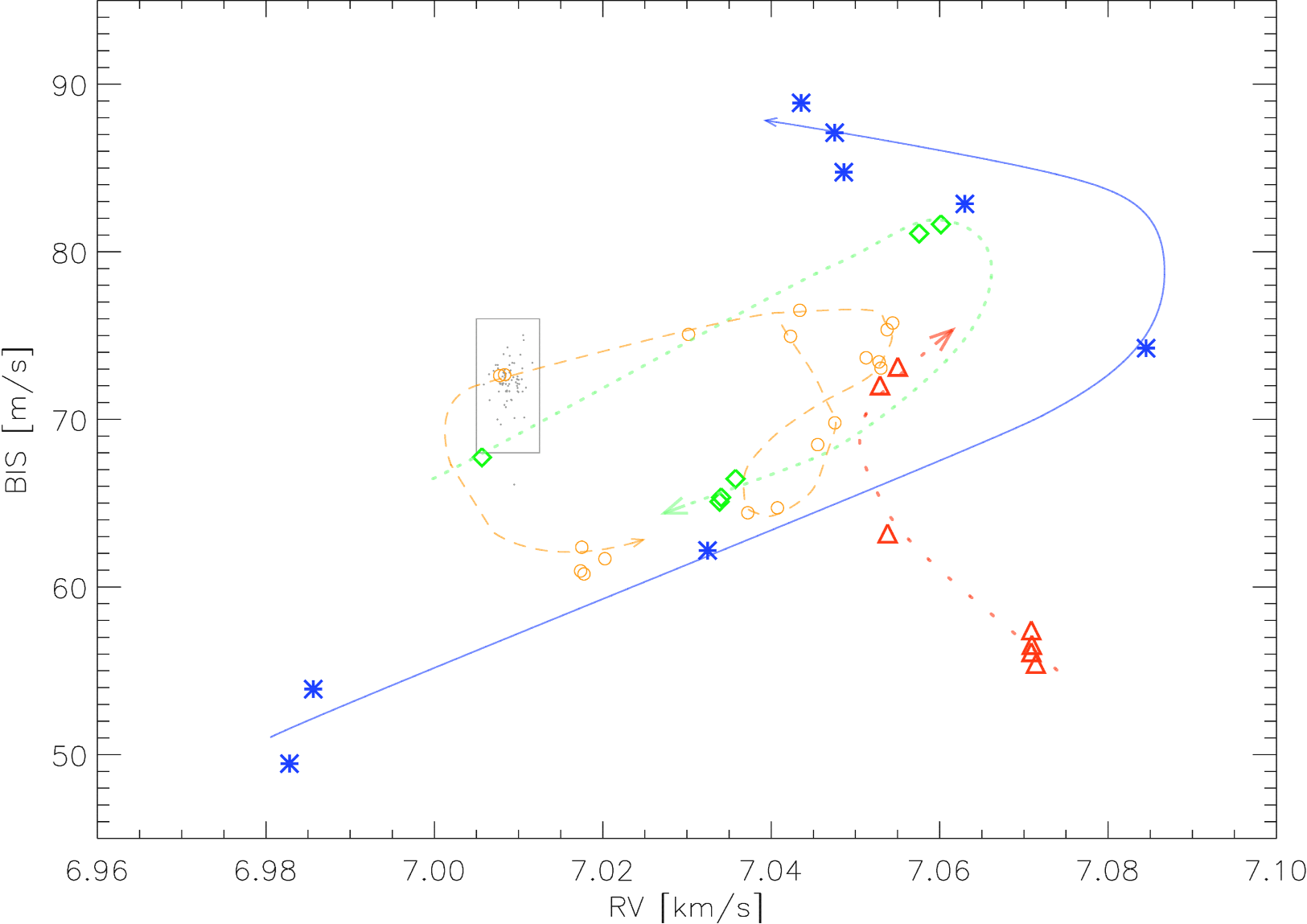}
\caption{\label{fig:bis} BIS vs.\ RV with the same symbols in Fig.~\ref{fig:rvbjd} and \ref{fig:rhk}.
The night-to-night variation is very small and comparable with the single-night scatter. 
Thus, the larger scale variation in BIS, which correlates well with RV, 
is real and attributable to line profile variations.
}
\end{figure}

It is well known that stellar magnetic activity affects the shape of spectral lines and thereby the apparent RV 
\citep[][]{gray1988,gray2005}, which again affects the ability to detect planets by the Doppler technique.
The best direct measure of activity available in the HARPS spectra are the emission cores of the 
calcium H and K lines, and for each spectrum we have derived the activity index $R_\mathrm{HK}$ following the 
procedure of Paper~I.
It was shown by  \citet{queloz+2001}
that the Bisector Inverse Slope (BIS) of the CCF
is a good qualitative measure of the 
distortion of the spectral lines caused by activity, and we have calculated the BIS
for all CCFs following the procedure of \citet{dall+2006}.  The series of RV, BIS and \rhk\ are shown
in Fig.~\ref{fig:rvbjd} along with the corresponding part of the photometric light curve.
In Figs.~\ref{fig:rhk} and~\ref{fig:bis} we show the correlations 
of RV with activity index $\log R_\mathrm{HK}$ and BIS.

From Figs.~\ref{fig:tsAll} and \ref{fig:rhk}, it is clear that the large scale RV variations 
are very well correlated with the activity as measured by \rhk.
As the star gets dimmer, it gets redder \citep[as noted by][]{strassmeier+1999} and the 
chromospheric activity gets higher, 
both facts pointing to cool spots with overlaid plages as the cause of the variation.
The variation of RV seems to be correlated with the photometric variation as well, although the correspondence is
not always straightforward and the RV amplitude varies differently from the light amplitude. 

The BIS shows in general
a strong correlation with RV as does \rhk. However, there seem to be significant
deviations: the red points ($\triangle$ symbols), corresponding to about two months in late 2006 only, show
a very distinct behavior in BIS, essentially reversing the sign of the RV-BIS correlation. 
This behavior is not seen in the direct activity measure $\log R_\mathrm{HK}$. 
The last observing period ($\circ$ symbols) exhibit variations in
RV and \rhk which is overall anti-correlated with the light variation, although much more complex. 
The BIS varies very little during this period, but like \rhk\ shows a quite complex behavior. 
This is in sharp contrast to the behavior of the blue points
($\ast$ symbols) where all parameters display large-amplitude, almost textbook-like behavior with
$\log R_\mathrm{HK}$ in anti-correlation with the light variation, and the BIS and RV being 
phase shifted with respect to the light.  Note that the internal precision on individual data points is very high
as evidenced by the low scatter of the three-night run.

It is also worth noting that the obvious strong variation of bisector shape is in apparent contradiction to the finding of
\citet{saar+donahue1997} who found that the BIS loses its diagnostic power for extremely low $v\sin i$. Their model results were confirmed
recently by \citet{desort+2007}, who found that a \vsini\ less than the spectrograph resolution (which is certainly the case for EK~Eri)
should yield a constant bisector shape.  This was confirmed by \citet{bonfils+2007} and \citet{huelamo+2008} who found
negligible BIS-RV correlations for the slow ($v\sin i < 1$~\kms) rotators \object{GJ 674} and \object{TW Hya}, respectively.
This is in contrast to our findings as evident from Fig.~\ref{fig:bis}. 
Presently, we can offer no explanation of this. 

Two other aspects are worth pointing out. First, while \citet{desort+2007} find the BIS-RV slope to be negative in all 
their simulations, we are clearly seeing mostly positive slopes, although with at least one period of 
negative slope as well.  These authors however presented only a few models and only for main sequence stars.
Secondly, while \citet{bonfils+2007} find a clear loop pattern in the plot of RV versus \ion{Ca}{ii} H+K emission, which they interpret
in terms of rotational modulation, our results are less clear, mostly due to the incomplete phase coverage.
However, from the time evolution of the \rhk--RV and BIS--RV relations as depicted in Figs.~\ref{fig:rhk} and~\ref{fig:bis}
we can see that there appears to be different segments of similar loop patterns for different observing seasons, which could
mean that these segments of the loops are not associated with the same spot regions. 
Indeed, if the star is viewed equator-on, then one spot or 
spot group is not enough to produce a sinusoidal light curve, much less the radial velocity and activity variations.

One possible conceptual model of EK~Eri is 
that of an oblique rotator seen close to $i=90^\circ$, with the magnetic axis 
of a dipolar field tilted with respect to the rotation axis. 
This assumes that we approximate the star as a single dipole,
which we believe is a reasonable approximation, taking into account also
the results of \citet{auriere+2008}. 
Assuming that the magnetic poles are associated with the spots,
we propose a model of EK~Eri involving two large spots or spot-covered areas, located 
180$^\circ$ opposite each other.
Although \citet{auriere+2008} observed no sign changes in the average longitudinal field over the
photometric phase $0.21 < \phi < 0.83$, which would correspond to the hypothetical two spots gradually
rotating into (respectively, out of) view, we note that their observations were performed during a period
where EK Eri seemed to be undergoing another cycle change similar to the one of 1987--1992. During these periods
the light amplitude is low, indicating that spots are small and possibly dominated by dynamo-related activity
rather than by the large scale dipole field.

In this scenario, the light curve minima correspond to a spot facing the observer, which happens twice per 
rotation. Hence, the
rotation period $P_\mathrm{rot}$ of the star is not equal to $P_\mathrm{phot}$, but rather twice that. 
This would mean that the true latitude-averaged rotation period of EK~Eri is $P_\mathrm{rot}=617.6$~d. 
The possibility of
$P_\mathrm{rot}$ being two or even three times the photometric period was also mentioned by \citet{auriere+2008} but
not investigated further. 
Such a relationship has been observed for the Sun \citep[e.g.,][]{durrant+schroter1983} where the rotation of active regions
across the solar disk gives rise to a 13-day period, i.e., half the solar rotation period. The phenomenon of
$P_\mathrm{rot} = 2P_\mathrm{obs}$ has also been suggested for Procyon \citep{arentoft+2008}.
Note that the period $2P_\mathrm{obs}$ would not show up in the period analysis unless the spots were 
of uneven size and stable over several rotations, which is clearly not the case.

 Although simple, this would explain the differences seen in the \rhk--RV loop patterns which seem to change direction
for every second loop, while also qualitatively explaining the appearance of the light curve. Of course, we are 
assuming that the field is poloidal, that the spots are associated with the large scale field, and that the star is seen 
equator-on. We have however not been able to produce a spot model that could reproduce the light curve satisfactorily
using this simple geometry, and the true structure of the magnetic field and the activity of EK~Eri is likely far more
complex.

\subsection{Photometric period changes}
As evident from the light curve (Fig.~\ref{fig:tsEph}) and as noted by \citet{strassmeier+1999}, the photometric period
of EK~Eri is not stable. In fact, based on the latest data, the star appears to be going through a phase similar to the 
period 1987--1992 where the light variations almost disappeared and the period changed significantly. 
While this may be explained by distinct magnetic cycles, 
other explanations have been proposed in view of the unusually long period.
One possibility proposed hypothetically by \citet{strassmeier+1990} was that the star is seen pole-on and that the long
period reflects activity cycles on a rapidly rotating star. Alternatively, they suggested a strong internal
rotation gradient to explain the activity in terms of a classical $\alpha\Omega$ dynamo. 
\citet{stepien1993} proposed the elegant solution that EK~Eri
is the descendant of a magnetic Ap star. Up to 10\% of Ap stars have rotation periods longer 
than 100~d \citep[e.g.,][]{mathys2008}, and a significant fraction of the field 
should be able to survive main sequence evolution \citep[][]{moss2003}.

\subsection{Magnetic field geometry, and low-amplitude stellar oscillations}
\citet{auriere+2008} observed modulations of the magnetic field strength with rotational phase, but they
were unable to relate it to the photometric period.
In particular, they observed a peak in the field strength at one phase, which they
assign as their phase zero point.   From our ephemeris given in Table~\ref{tab:results} we find that their
zero phase corresponds to $\phi_\mathrm{phot}=0.21$ in our data i.e., very close to quarter photometric phase.

In our RV monitoring, we have covered the interval from time of photometric minimum ($\phi_\mathrm{phot}=0$) to first quarter
phase on two segments of the light curve; in 2004--5 and 2006--7 (see Fig.~\ref{fig:tsAll}). 
In the first segment, all parameters vary smoothly. As we noted, this is the classical textbook behavior. 
In the second segment, the activity
level as inferred by $\log R_\mathrm{HK}$ is overall higher, with the maximum level likely to have been reached 
just before 2006-11-06, but after the time of light minimum, i.e. around $\phi_\mathrm{phot}=0.1$--$0.2$.
It is likely then, that the high field strength measured by \citet{auriere+2008} corresponds to a similar
high value of $\log R_\mathrm{HK}$ and vice versa. Furthermore, around $\phi_\mathrm{phot}=0.2$ in 2006--7, 
the BIS changes direction
for a short while,
which may indicate an additional contribution on top of the main spot (group), possibly a smaller bright spot, or
the emergence of a short lived spot. If an $\alpha^2$ dynamo can work locally to augment a large scale seed field
at small scales on the stellar disk 
\citep[a mechanism proposed for AGB stars by][]{soker+zoabi2002},
then this spike could be caused by spots with completely different characteristics in terms of size and temperature, 
appearing alongside the semi-static fossil field-induced spots responsible for the main light modulation.

The magnetic field geometry suggested by \citet{auriere+2008} is close to a dipole, which makes intuitive sense for
a remnant static field. One may expect a large scale magnetic field to have a stabilizing effect on the overall shape
of the star, i.e., resisting deformations caused by the oscillations.  As noted by \citet{braithwaite2009}, a dominantly poloidal
field tends to align the magnetic axis perpendicular to the rotation axis, thereby contributing to making the star
oblate, even for very slow rotation.
With this in mind, the surprisingly low oscillation 
amplitude-per-radial-mode (cf.\ Sect.~\ref{sec:osc})
may have a logical explanation. In most stars, the $l=1$ modes are expected to have the highest
amplitudes, but if the low-degree modes are suppressed, then the bulk of the oscillations will be
higher degree modes.  Assuming an expected amplitude per mode of $\approx 0.5$~\ms\ \citep[Fig.~8 of][]{kjeldsen+2008}
we find a spatial response scaling factor $S_l/S_0 = 0.3$. Comparing with Table~1 of \citet{kjeldsen+2008} we then
propose that the bulk of the oscillations in EK~Eri could be higher degree modes of $l=3, 4$. 
Coincidentially, for the rapidly oscillating magnetic Ap stars (roAp), which are possible progenitors of EK~Eri,
there is emerging evidence for interaction between the magnetic field and the oscillation driving mechanism
\citep[e.g.,][and references therein]{saio+2010}.

Of course our time coverage is very poor and the discrepancy may very well be due to mode beating and the stochastic
nature of mode excitation. 
Obviously, longer observing runs are required to test this scenario. 
It is interesting to note that the magnetic field measurement of \citet{auriere+2008} as well as our
asteroseismic results were obtained while the star was apparently entering another ``bright'' phase akin to the 1987--1992
period. In order to test the interaction between the magnetic field and the oscillations it would be highly
desirable to conduct asteroseismic measurements in periods of high field strength as well as in periods of relatively
low field strength.

\section{Conclusions}
\label{conclusions}
We have presented results from an intensive monitoring of the active sub-giant star EK~Eri.
We have used photometric data covering 30 years and 
spectroscopic data probing long-term variation in activity during 3 years and 
high-cadence radial-velocity monitoring from 3 nights. 
From the photometry we have refined the rotation period and the ephemeris as listed in Table~\ref{tab:results}.
Also, from 3 half-nights of high-cadence RV monitoring we have detected solar-like oscillations
in a late-type spotted sub-giant star for the first time. 
While oscillations have been detected in a number of late-type giants
\citep[e.g.,][]{hekker+2009,hatzes+zechmeister2008} 
and for mildly active solar-type stars \citep[e.g., Pollux:][]{hatzes+zechmeister2007,auriere+2009}, 
this is the first detection for a sub-giant hosting a strong magnetic field.
Unfortunately our data do not allow us to resolve individual frequencies.
We measure an amplitude per radial mode of $\approx 0.15$~\ms\ at a position of maximum power $\nu_\mathrm{max}=320\pm32$~$\mu$Hz.
This amplitude is at least a factor of 3 lower than expected and, if confirmed, may mean that
the magnetic field has a strong
stabilizing effect on the stellar geometry, essentially favoring high-degree oscillation modes in the presence of
a near-dipole magnetic field. 
In that case, the interpretation of the asteroseismic data may become 
more difficult for this and other sub-giant stars with similar magnetic properties. 
A longer time series is required in order to obtain quantitative results.
For reference purposes, having accurate oscillation data for a non-active star
at the position in the HR diagram of EK~Eri would be highly desirable. Unfortunately, neither
CoRoT nor Kepler apparently covers this.

Based on the roughly sinusoidal shape of the light curve, the likely very high inclination, 
the field geometry suggested by \citet{auriere+2008}, 
and the behavior of the activity indicators as function of RV, 
we suggest a conceptual model of EK~Eri with two large 
low-latitude spot covered areas approximately $180^\circ$ apart on a star viewed equator-on. 
In this scenario, the rotational period is twice the photometric period, 
thus $P_\mathrm{rot} = 2P_\mathrm{phot} = 617.6$~d.
We note however, that a simple two-spot model is not able to account for all the seasonal light variations 
observed, mostly due to the unknown spot lifetimes, sizes and longitudes.

Regardless of the rotation period, the measured values of $v\sin i$ both from this work and from the literature are
inconsistent with the derived radius of the star. Both the radius derived from asteroseismology and from the 
spectral analysis set strict upper limits on $v\sin i$ which are lower than previous estimates.
We thus conclude that the $v\sin i$ is too low to be reliably measured with available spectrographs.

Based on high-quality HARPS spectra we have derived the atmospheric parameters of EK~Eri to very high
precision. The abundance pattern for 17 analysed elements is very similar to the Sun,
and we detect no anomalies that could otherwise be attributed
to an earlier evolutionary state as a magnetic Ap star.
However, in order to argue
for or against EK~Eri being a descendant of a magnetic Ap star, stronger
constraints on the mass and evolutionary state are needed. Further seismic studies,
preferably at varying rotational phases, may deliver such constraints
in terms of accurate asteroseismic mass and radius measurements,
and are also needed to probe the possible link between solar-like oscillations and the magnetic field.

\begin{acknowledgements}
DS would like to acknowledge support from the Australian Research Council.
TD acknowledges support by the Gemini Observatory, which is operated by the Association of Universities 
for Research in Astronomy, Inc., on behalf of the international Gemini partnership of Argentina, 
Australia, Brazil, Canada, Chile, the United Kingdom, and the United States of America.
This research has made use of the SIMBAD database, operated at CDS, Strasbourg, France.
We would like to thank the anonymous referee for comments and suggestions that have improved the paper.
\end{acknowledgements}

\bibliographystyle{bibtex/aa}
\bibliography{13710_dall}

\end{document}